\documentstyle[aps]{revtex}
\begin{document}
\title{Matrices on a point as the theory of everything}
\author{Vipul Periwal}
\address{Department of Physics,
Princeton University,
Princeton, New Jersey 08544}

\def\dd{\hbox{d}}
\def\tr{\hbox{tr}}\def\Tr{\hbox{Tr}}
\maketitle
\begin{abstract}It is shown that the world-line can be eliminated in
the matrix quantum mechanics 
conjectured  by Banks, Fischler, Shenker 
and Susskind to describe the light-cone physics of M theory.  The 
resulting matrix model has a form that suggests origins in the  
reduction to a point of a Yang-Mills theory.  The reduction of the
Nishino-Sezgin $10+2$ dimensional supersymmetric Yang-Mills theory
to a point gives a matrix model with the appropriate features: Lorentz
invariance in $9+1$ dimensions, supersymmetry, and the correct number 
of physical degrees of freedom.\end{abstract} 

\def\al{\alpha}
\def\be{\beta}
Banks, Fischler, Shenker and Susskind\cite{banks} have conjectured that
M theory, in the light-cone frame, is exactly described by the large 
$N$ quantum mechanics of a particular supersymmetric matrix model, a
relation suggested first by Townsend\cite{town}.
The concrete motivations for this conjecture are the work of 
Witten\cite{ed} on the dynamics of D-branes, and the work of de Wit,
Hoppe and Nicolai\cite{dewit} on a discretization of the supermembrane
action.  Supporting evidence for this conjecture has been given by
Berkooz and Douglas\cite{berk}.

The aim of the present note is to show that the model of \cite{banks}
can be re-written as just a matrix model, in other words, 
the world-line of the quantum mechanics matrix model can be 
eliminated entirely.  The form that I find of this 
matrix model suggests that there is a particular Lorentz-covariant 
matrix model that underlies it---the reduction to a point ({\it 
i.e.}, $0+0$ dimensions) of the
$10+2$ dimensional super Yang-Mills equations and symmetries found by
Nishino and Sezgin\cite{nish}.  While the symmetries and  the 
degrees of freedom of the model provide evidence that this model underlies the 
model of Banks et al., I have not been able to show this directly.
Since the light-cone model does not
include all the physics of M theory\cite{banks}, nor is light-cone quantization 
with periodic boundary conditions without subtleties, this failure may not  
be a flaw in the covariant model presented in this note.  (The $10+2$-
dimensional model reduces directly to $9+1$-dimensional super 
Yang-Mills\cite{nish}, and the further reduction to $0+0$ of this theory appears 
to agree with $S_{l-c}$, but without the light-cone interpretation 
given in \cite{banks}.)

One motivation for what follows is the observation that the 
matrix character of the quantum mechanics arose in \cite{dewit} from
a discretization of the membrane volume, after an identification of
the time on the membrane world-volume with a space-time light-cone
coordinate.  This construction should be more symmetric both from the 
world-volume diffeomorphism point of view, and from the point of view
of space-time Lorentz invariance.  What follows is a concrete 
realization of this `symmetrization'.

The bosonic part of the action considered by Banks et al.\cite{banks} is
\begin{eqnarray*}
 S_{l-c} = \int \dd t \ \tr \left( \dot {X^{i}}\dot {X^{i}} - {1\over 2} 
 \left[X^{i},X^{j}\right]\left[X^{i},X^{j}\right] \right).
\end{eqnarray*}
Here $X^{i}$ are Hermitian $N\times N$ matrices and $i=1,\dots, 9,$
with repeated indices summed.  This model is obtained by the 
dimensional reduction of $N=1$ super Yang-Mills theory in $D=10$ to
the world-line of a $0-$brane.  Following \cite{dewit}, a large $N$
limit of this model can be identified with the supermembrane action in
light-cone gauge via a map from the generators of the SU$(N)$ Lie
algebra into the generators of area-preserving diffeomorphisms on
a spherical membrane.  See also \cite{barstwo}.

To start, let us discretize the world-line, so that it consists of a 
set of points with a spacing $\epsilon.$  Then $\int \dd t$ goes over 
into $\sum_{j=-\infty}^{j=+\infty} \epsilon,$ and the derivatives
turn into $(X(j+1)-X(j))/\epsilon.$  The powers of $\epsilon$
that occur in $S_{{l-c}}$ can be absorbed into a rescaling of $X^{i},$
so we set $\epsilon=1.$
Now define block diagonal matrices $Y^{i}$ such that $X^{i}(j)$ occurs 
as the $j^{{\rm th}}$ block along the diagonal.  Let $Y^{+}$ be the matrix 
that represents the shift operator (on blocks of length $N$), 
and $Y^{-}$ its adjoint.  Then, with $\Tr \equiv \tr_{Y}=\tr_{X}\sum$ 
in an obvious notation, 
\begin{eqnarray*}
S_{l-c}= {1\over2}\Tr [Y^{\alpha},Y^{\beta}][Y_{\alpha},Y_{\beta}],
\end{eqnarray*}
with $Y^{\pm}\equiv Y_{{\mp}},$ and $\al,\be = +,-,1,\dots, 9.$
The complete supersymmetric action can also be written in 
this form.  There is really no need to go through the discretization
explicitly of course, since we are just tensoring the algebra of 
functions on the world-line with a matrix algebra, giving a bigger `matrix'
algebra.  The introduction of two matrices (which are, of course,
adjoints) is also unnecessary if one
does not discretize the world-line.

Thus, as claimed, I have shown that the matrix quantum mechanics 
model is obtained as a particular case of a matrix model with two 
{\it more} matrices, with the additional matrices of a fixed 
non-dynamical form.  The simple form obtained suggests
that a reduction of a supersymmetric Yang-Mills theory down to 
$0+0$ dimensions might be the underlying gauge-invariant and 
Lorentz-invariant system, since such a reduction would give a bosonic
term with exactly such a trace of 
the square of a commutator. The question that remains is: What 
supersymmetric Yang-Mills theory would give rise to an appropriate
model?

Motivated by the work of Blencowe and Duff\cite{blen}, 
Vafa\cite{vafa}, Hull\cite{hull}, Kutasov and Martinec\cite{kut},
and Bars\cite{bars}, Nishino and Sezgin\cite{nish} have given an
elegant construction of a supersymmetric Yang-Mills 
model in $10+2$ dimensions\cite{van} that has the
following features: 
({1}) It features constraints on the field strength and the 
fermion that are based on the choice of  a constant null vector, 
reducing the invariance group to the little group of the null vector.
(2) Besides the usual gauge symmetry,
it has one additional bosonic gauge symmetry.
(3) Ordinary dimensional reduction to ten dimensions leads
to the usual supersymmetric Yang-Mills equations.
(4) No action is known for this model, just the equations of motion and
the symmetries.
\par\noindent As mentioned earlier, one could have started with the 
ten-dimensional theory, but the twelve-dimensional equations seem to
be more general, thus may be required for certain F-theory 
purposes(\cite{vafa,kut,hull,bars}).

\def\sig{\sigma}
\def\la{\lambda}
\def\ga{\gamma}
Nishino and Sezgin\cite{nish} consider a vector field and a 
positive chirality Majorana-Weyl fermion 
in $10+2$ dimensions. Dimensionally reduced to $0+0$ dimensions, 
their equations of motion\cite{nish} are ($F_{\mu\nu} 
\equiv[A_{\mu},A_{\nu}]$)
\begin{eqnarray*}
[A_{\mu},F^{\mu}{}_{[\rho}]n_{\sig\vphantom{\rho}]} + {1\over 
4}\{\bar\la,\ga_{{\rho\sig}}\la\} = 0, \qquad 
\ga^{\mu}[A_{\mu},\la]=0,
\end{eqnarray*}
with the constraints
\begin{eqnarray*}
n^{\mu}[A_{\mu},A_{\nu}]=0,\qquad n^{\mu}[A_{\mu},\la]=0,\qquad 
\hbox{and} \qquad n^{\mu}\ga_{\mu}\la=0.\end{eqnarray*}

The supersymmetry transformations are 
\begin{eqnarray*}\delta_{Q}A_{\mu}=\bar\epsilon\ga_{\mu}\lambda, \qquad 
\delta_{Q}\la = 
{1\over4}\ga^{{\mu\nu\rho}}\epsilon[A_{\mu},A_{\nu}]n_{\rho}.\end{eqnarray*}
Besides the usual gauge transformations, there is a new local gauge 
transformation 
\begin{eqnarray*}
\delta_{\Omega} A_{\mu}\equiv\Omega n_{\mu}, \qquad  
\hbox{with}\qquad \Omega: 
[\Omega,n_{\mu} A^{\mu}]=0.\end{eqnarray*} 
The commutator of supersymmetry transformations gives 
\begin{eqnarray*}
 [\delta_{Q}(\epsilon_{1}),\delta_{{Q}}(\epsilon_{2})]=
\delta_{\xi}+\delta_{\Lambda}+\delta_{\Omega},\end{eqnarray*}
where $\delta_{\xi}$ is a translation by 
$\xi\equiv\bar\epsilon_{2}{\ga^{\mu\nu}}\epsilon_{1} n_{\nu},$
$\delta_{\Lambda}$ is a gauge transformation by 
$\Lambda=-\xi^{\mu}A_{\mu},$ and
$\Omega= \hbox{${1\over 2}$}F_{\mu\nu}\bar \epsilon_{2}\ga^{{\mu\nu} }
\epsilon_{1}.$  
In the dimensionally reduced model, the translation is 
set to zero, and the gauge transformation by a field dependent 
parameter is the analogue of translation.  Thus, 
the supersymmetry algebra on $\Omega$-gauge-invariant states is
\begin{eqnarray*}
\{Q_{\al},Q_{\be}\}= \ga^{\mu\nu}_{\al\be}n_{\nu}P_{\mu}.\end{eqnarray*}
This algebra can be found in \cite{bars}.
An important consequence of this identification of $P_{\mu}$ is that
observables that are $\Lambda$-gauge-invariant, are automatically
translation invariant, quite appropriate for physical observables in
a theory of quantum gravity.

Counting degrees of freedom, starting from 12 matrix degrees of 
freedom, we see that there are 9 matrix degrees 
of freedom, since there is one constraint, and there are two gauge
symmetries.  However, the gauge symmetries and constraint do not
remove {\it all} the degrees of freedom of three matrices.  The adjoint
representation action of the $\Lambda$ gauge symmetry can be used to 
reduce one of the matrices to a diagonal form.  The remaining shift
symmetry has a parameter that is constrained, and hence again does not
suffice to eliminate all the degrees of freedom of a matrix.  This is
precisely the general structure we must obtain if this model is to 
describe the same physics as $S_{{l-c}}.$  It is these residual 
degrees of freedom that are interpreted as giving rise to dynamics in 
this Lorentz-invariant model.  

In summary, I showed that the light-cone matrix quantum mechanics action, 
$S_{l-c},$ can be written as a matrix model action with 9 matrix 
degrees of freedom and two additional matrices of fixed form, in a 
form similar to  the dimensional reduction of a
Yang-Mills action to a point, {\it i.e.} to $0+0$ dimensions.  This
rewriting can be related to the ideas of Connes\cite{connes}, but the physics 
that follows is 
not greatly illuminated by making such a connection.
Motivated by this, I observed that the Nishino-Sezgin $10+2$ 
dimensional supersymmetric Yang-Mills equations, reduced to a point,
have exactly the 
symmetries and degrees of freedom appropriate for a Lorentz invariant
(in $9+1$ dimensions) supersymmetric matrix model underlying $S_{l-c}.$
The $A_{\mu}$ matrices are identified with translation operators, on 
account of the fermion equation of motion, and the form of the
supersymmetry algebra.  A new feature here is the natural 
interpretation of gauge transformations, when reduced to 0$+$0
dimensions, as translation operators.
The equations of the model are expressed in terms of these
translation operators, which may be related to Witten's original 
interpretation\cite{ed} by $T$-duality.    
The Lorentz group is the little group of the null vector in $10+2$
dimensions that appears in the defining constraints.  While there are
remnants of the full $10+2$-dimensional Lorentz invariance in the 
equations, it is not clear if there is any limit of the model in which
$10+1$-dimensional Lorentz invariance is exactly realized.  There
is, of course, 
no physical reason to suppose that there should be an uncompactified
$10+1$-dimensional theory with such an invariance.

The operator equations of motion should provide a complete definition
of the theory, due to the supersymmetry.  
Observables are $\Lambda$- and $\Omega$-gauge 
invariant quantities, constructed from matrices that satisfy these
equations of motion.  
The emergence and interpretation of dynamics 
depends on the solution of the constraints, the separation of 
gauge degrees of freedom, and on what is treated as `background'
geometry.  This is entirely appropriate for a theory 
of quantum gravity, based on the uncertainty principle, as
embodied in the non-commuting translation operators $A_{\mu},$
the principle of equivalence, as embodied in the Lorentz invariance at
a point, and supersymmetry.   
\acknowledgements

I am grateful to the D-brane journal club for valuable explanations.
This work was supported in part by NSF grant PHY96-00258.

\end{document}